\documentstyle[12pt]{article}
\begin{document}

\def\lsim{\mathrel{\rlap{\lower3pt\hbox{\hskip0pt$\sim$}}
    \raise1pt\hbox{$<$}}}         
\def\gsim{\mathrel{\rlap{\lower4pt\hbox{\hskip1pt$\sim$}}
    \raise1pt\hbox{$>$}}}         

\newcommand{\beq}{\begin{equation}}
\newcommand{\eeq}{\end{equation}}
\newcommand{\aver}[1]{\langle #1\rangle}

\newcommand{\La}{\overline{\Lambda}}
\newcommand{\Lam}{\Lambda_{\rm QCD}}

\newcommand{\lhs}{{l.h.s.} }
\newcommand{\rhs}{{r.h.s.} }

\newcommand{\ind}[1]{_{\begin{small}\mbox{#1}\end{small}}}
\newcommand{\hscale}{\mu\ind{hadr}}

\newcommand{\appa}{\mbox{\ae}}
\newcommand{\al}{\alpha}
\newcommand{\as}{\alpha_s}
\newcommand{\GeV}{\,\mbox{GeV}}
\newcommand{\MeV}{\,\mbox{MeV}}
\newcommand{\matel}[3]{\langle #1|#2|#3\rangle}
\newcommand{\state}[1]{|#1\rangle}
\newcommand{\ra}{\rightarrow}
\newcommand{\ve}[1]{\vec{\bf #1}}

\newcommand{\eq}[1]{eq.\hspace*{.15em}(\ref{#1})\hspace*{-.3em} }
\newcommand{\eqs}[1]{eqs.\ \hspace*{-.15em}(\ref{#1})}

\newcommand{\re}[1]{Ref.~\cite{#1}}
\newcommand{\res}[1]{Refs.~\cite{#1}}

\begin{titlepage}
\renewcommand{\thefootnote}{\fnsymbol{footnote}}

\begin{flushright}
CERN-TH/96-298\\
UND-HEP-96-BIG\hspace*{.15em}$03$\\
hep-ph/9611259 \\
\end{flushright}

\Large
\begin{center} 
{\bf Theoretical Aspects of \\ the Heavy Quark Expansion}\\
\vspace*{1.4cm}
\large
Nikolai Uraltsev
\vspace*{.4cm}\\
\normalsize 
{\it TH Division, CERN, CH 1211 Geneva 23, Switzerland},\\
{\it Dept. of Physics, Univ. of Notre Dame du Lac, Notre Dame, 
IN 46556, U.S.A.
}\\
and\\
{\it Petersburg Nuclear Physics Institute,
Gatchina, St.Petersburg 188350, Russia$^{\,\dagger}$
\vspace*{1.9cm}}\\
{\large\bf Abstract} \vspace*{.02cm}
\end{center}

\normalsize
I give a brief outline of the theoretical framework for the modern 
treatment of the strong interaction effects in heavy quark decays, 
based 
on first principles of QCD. This model-independent approach is required to
meet the precision of current and future experiments.
Applications to a few problems of particular practical interest are reviewed,
including the precise determination of $V_{cb}$ and $V_{ub}$. I emphasize
the peculiarities of simultaneously accounting for the perturbative and
power-suppressed effects necessary for accurate predictions. 
\vfill

\begin{center}
Invited lecture presented at the Workshop \\{\it Beauty `96}\\
Rome, 17--21 June 1996 
\end{center}
\vfill
\nopagebreak

\begin{flushleft}
CERN--TH/96--298\\
October 1996\\
\rule{2.4in}{.25mm} \\
$^\dagger$ Permanent address.\\

\end{flushleft}
\end{titlepage}

\newpage

\section{Introduction}

A key role in exploring the Standard Model is played by
studying electroweak heavy flavor decays. It was realized 20 years ago  
that the strong interaction effects in heavy flavor hadrons can be
treated within QCD. Yet the full power of theoretical methods
acquired in QCD was applied here only recently. 
They were developed along two main directions, `symmetry-based' and 
`dynamical'. These two
lines in the heavy quark theory were the counterparts of the basic
theoretical strategy in studying strong interactions: {\em isotopic 
invariance} and {\em chiral symmetry} on the one hand, and {\em asymptotic
freedom} on the other.

In heavy quark physics, the early period to the end of the 80s saw
mostly the dynamical approach applied at a simplified
`intuitive' level. The nonperturbative effects were often thought to be small
even in the decays of charm particles. 
The following few years were dominated by `symmetry' considerations; 
the operating language for those analyses was the so-called 
Heavy Quark Effective Theory (HQET), which incorporated some basic elements 
of the 
general heavy quark expansion in QCD (HQE) but was limited only to certain 
classes of processes.

Finally, over the last few years a consistent well-defined dynamical
approach has been developed, which automatically respects the heavy quark
symmetries in a manifest way. 
Here the most precise determinations of $|V_{cb}|$ and $|V_{ub}|$ were 
made.

The main
effects in a weak decay of heavy quarks $Q$ originate from distances 
$\sim 1/m_Q \ll 1/\Lam$. Since $\as(m_Q)\ll 1$ they are tractable through 
perturbation theory. The QCD interaction becomes strong only when the momentum
transfer is much smaller than the heavy quark mass, $k \ll m_Q$. Two basic 
ingredients of HQE are thus elucidated:\\
$\bullet$ The {\em nonrelativistic expansion}, which yields the effects of 
`soft' physics in the form of a power series in $1/m_Q$.\\
$\bullet$ The treatment of the 
strong interaction domain based on the 
{\em Operator Product Expansion} (OPE).\\
Unless an analytic solution of  QCD is at hand, these two elements appear to be
indispensable for heavy quark theory. 

The general idea of separating the two 
domains and applying different theoretical tools to them was formulated long
ago by K.~Wilson \cite{wilson} 
in the context of problems in statistical mechanics; in the
modern language, applied to QCD it is similar to lattice gauge theories. 
The novel feature we face in the theoretical analysis of beauty decays
is that they often allow -- and even demand by virtue of the 
existence of
precise experimental measurements -- rather accurate predictions,
requiring a {\em simultaneous} treatment of perturbative and
nonperturbative QCD effects with enough precision in both. This problem is
not new; the theoretical framework has been elaborated more than 10 years ago
\cite{fail}, but its phenomenological implementation was not mandatory 
until recently. 
Failure to incorporate it properly leads to certain theoretical
paradoxes and, unfortunately, some superficial controversy in the numerical
estimates in the literature.

With significant progress made over the last years, the theory of
the heavy flavors is still not a completed field and is undergoing to further 
extensive development.  
I will focus on a few selected
topics that illustrate the theoretical framework, 
and review 
the overall status
of the heavy quark expansion,  with the main emphasis on
the qualitative features. Some important theoretical 
applications are presented in the lectures by C.~Sachrajda \cite{sach96}
(exclusive decays), A.~Ali \cite{ali96} (rare $b$ decays) and M.~Gronau 
\cite{gron96} 
($CP$ violation) (these Proceedings). Additional theoretical aspects are 
covered
in the summarizing contribution by G.~Martinelli \cite{mart96}, where 
recent experimental data are also discussed.

\section{Semileptonic decays}

\begin{figure}
\vspace{2.8cm}
\includegraphics{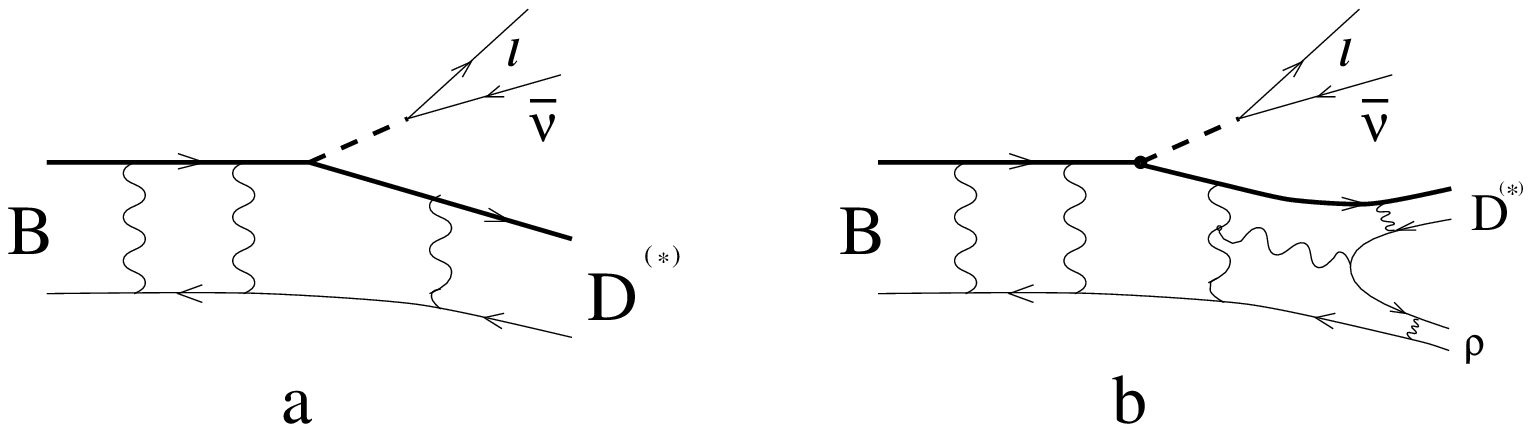}
\caption{
Exclusive $B\ra D^{(*)}$ ({\bf a})\hspace*{.5em} and generic
({\bf b})\hspace*{.5em} semileptonic decays.}
\end{figure}

The QCD-based heavy quark expansion can equally be applied 
to all types of heavy flavor transitions. Semileptonic decays are the
simplest case and I shall devote most of the attention to them; for practical
reasons I focus on $b\ra c$ transitions. A brief discussion of the $b\ra u$
decays will be given later.

A typical semileptonic decay is schematically shown in Figs.~1. Generally, two
types of decay rates can be singled out: inclusive widths where any combination
of hadrons is allowed in the final state, and exclusive decays, when a
transition into a particular charmed hadron is considered, usually $D$ or
$D^*$.

\subsection{Inclusive semileptonic width}

The semileptonic width of a heavy quark has the form
\beq
\Gamma_{\rm sl} = \frac{G_F^2 m_b^5}{192\pi^3} |V_{cb}|^2 \cdot z_0
\left(\frac{m_c^2}{m_b^2}\right) \cdot \appa\;\;,
\label{1}
\eeq
where $z_0$ is the known phase space suppression factor and $\appa$ generically
includes all QCD corrections. In the heavy quark limit the difference resulting
from using the quark mass $m_b$ and the meson mass $M_B$ in \eq{1} disappears:
\beq
(M_B-m_b)/m_b\;\sim \; 1/m_b\;\;.
\label{2}
\eeq
For the actual $b$ quark $M_B^5/m_b^5$ amounts to a
factor of $1.5$--$2$, which formally constitutes a power-suppressed 
effect. This demonstrates the necessity of a systematic control of
nonperturbative corrections even in decays of beauty particles.

The central result obtained by direct application of OPE to the inclusive decay 
widths in QCD is the absence of $1/m_Q$ corrections \cite{buv} -- in contrast
with the presence of such terms in the hadron 
masses. 
The physical reason behind this fact is the
conservation of the color flow in QCD, which leads to the cancellation of the
effects of the color charge (Coulomb)  interaction in the initial and final
states. In terms of 
nonrelativistic quantum mechanics (QM), it is the 
cancellation between the phase space suppression caused by the Coulomb binding
energy in the initial state, and the Coulomb distortion of the final state
quark wavefunctions. The inclusive nature of the total widths ensures that 
they are sensitive only to the interaction on the time scale $\sim
1/\Delta E \sim 1/m_b$.
The final-state-interaction effect is thus not determined by 
the actual behavior of 
the strong forces at large distances, but only by the
potential in the close vicinity of the heavy quark. The cancellation 
therefore occurs universally, whether or not a nonrelativistic QM 
description is applicable.

The leading power corrections start with terms $1/m_b^2$; they were 
calculated in \cite{buv} and are expressed in terms of the expectation values
of two operators of dimension $5$, which have a transparent QM interpretation:
\beq
\mu_G^2\simeq \frac{1}{2M_B}\matel{B}{\bar b \frac{i}{2}
\sigma_{\mu\nu}G^{\mu\nu} b}{B} \leftrightarrow \matel{B}{\vec{\sigma}_b
\cdot g_s\vec{{\cal H}_g}}{B} \simeq \frac{3}{4} (M_{B^*}^2-M_B^2)\simeq 0.35
\GeV^2
\label{4}
\eeq
\beq
\mu_\pi^2\simeq \frac{1}{2M_B}\matel{B}{\bar b (i\vec{D}\,)^2b}{B} 
\leftrightarrow \matel{B}{\vec{p}^{\,2}}{B} \;\;.
\label{5}
\eeq
The value of $\mu_\pi^2$ is not yet known directly; a model-independent 
lower bound was established in \cite{vcb,volkin}:
$\;\mu_\pi^2\: > \: \mu_G^2\:$;
this puts an essential constraint on its possible values. This bound is in 
agreement with QCD sum rule calculations \cite{pp}, 
yielding a value of about $0.5\GeV^2$ 
and with a more phenomenological estimate \cite{third}.
In the absence of gluon corrections, as in 
simple QM models, the expectation value $\mu_\pi^2$
would coincide with the HQET parameter $-\lambda_1$; they are different,
however, in the actual field theory, where both $\mu_\pi^2$ and $\mu_G^2$ 
depend on the normalization point. 

Including the nonperturbative corrections, the semileptonic width has the
following form \cite{buv,bs,bbsuv,prl}:
\beq
\Gamma_{\rm sl} = \frac{G_F^2 m_b^5}{192\pi^3} |V_{cb}|^2 \left\{z_0
\left(1-\frac{\mu_\pi^2-\mu_G^2}{2m_b^2} \right)
-2\left(1-\frac{m_c^2}{m_b^2} \right)^4\frac{\mu_G^2}{m_b^2}
-\frac{2}{3} \frac{\as}{\pi} z_0^{(1)} + ...
\right\}\;.
\label{7}
\eeq
The $1/m_b^2$ corrections to $\Gamma_{\rm
sl}$ are rather small, about $-5\%$, and increase the
value of $|V_{cb}|$ by $2.5\%$; the impact of the higher order power
corrections is negligible.

Good control of the QCD effects in the inclusive semileptonic widths provides
the most 
accurate direct way to determine $|V_{cb}|$ in a truly model-independent 
way. It sometimes faces a traditional scepticism: which numerical value
must 
be used for $m_b$ and $m_c$? This practical problem has
deep roots; failure to understand them is the major source of
controversy about masses and inclusive widths found in the literature. It
will be briefly discussed below. 
In reality, the precise value of $m_b$ is not too important, since the $b\ra c$
width depends to a large extent on the difference $m_b-m_c$ rather than on
$m_b$ itself; the former is constrained in the HQE:
\beq
m_b-m_c=\frac{M_B+3M_{B^*}}{4}-\frac{M_D+3M_{D^*}}{4} + \mu_\pi^2
\left(\frac{1}{2m_c}-\frac{1}{2m_b}\right) + ... \approx  3.50\GeV\;.
\label{8}
\eeq
It also independently enters lepton spectra in semileptonic decays \cite{prl}
and can be extracted from the data \cite{volspec}. Numerically 
\cite{vcb,upset}, a change in $m_b$ by $50\MeV$ leads 
only to a $1\%$ shift in $|V_{cb}|$.\vspace*{.3cm}

\noindent
{\it Heavy quark masses}\vspace*{.2cm}

\noindent
The controversy about $m_b$ is due to the fact that HQET 
was popularly based on the
so-called `pole' mass of the heavy quarks. Not only was it a starting
parameter of the HQET-based expansions, it is this pole mass that one always 
attempted to extract from the experimental data. It
turns out, however, that the pole mass of the heavy quark is not a 
direct observable and its definition suffers from an irreducible intrinsic 
theoretical uncertainty of order $\Lam$ \cite{gurman}.  

At first sight this looks paradoxical and counter-intuitive: 
for example, the value of $m_e$ quoted in the
tables of physical constants is just the {\em pole} mass of the electron. 
In QCD there is no `free heavy quark' particle in
the physical spectrum, and its pole mass is not well defined. 
The problems facing the possibilities to extract the pole mass from typical 
measurements were illustrated in Refs.~\cite{optical} and \cite{bsg}.

\begin{figure}
\vspace{2.8cm}
\includegraphics{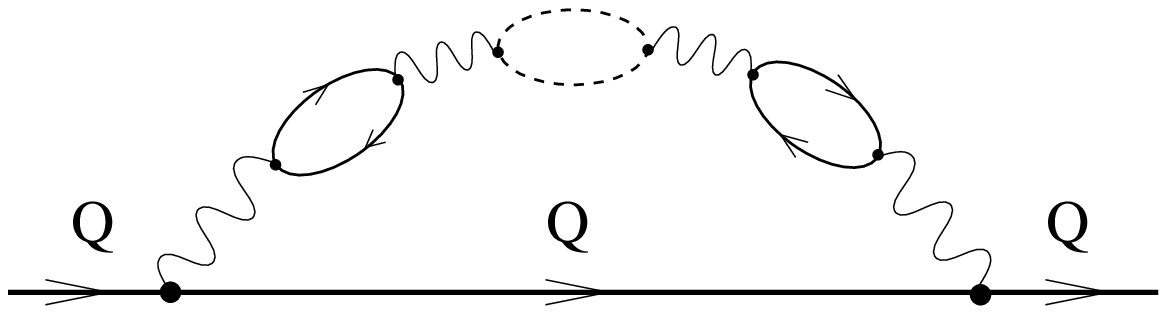}
\caption{
Perturbative diagrams leading to the IR renormalon
uncertainty in $m_Q^{\rm pole}$ of the order of $\Lam$. The contribution of
the gluon momenta below $m_Q$ expresses the classical Coulomb self-energy of
the colored particle. The number of bubble insertions into the gluon propagator
can be arbitrary.}
\end{figure}

The physical origin of the uncertainty $\delta m_Q^{\rm pole} \sim \Lam$ is the
gluon Coulomb self-energy of the 
static colored particle. The energy stored in the
chromoelectric field inside a sphere of radius $R \gg 1/m_Q$ is given by 
\beq
\delta E_{\rm Coulomb} (R) \propto 
\int_{ 1/m_b \sim |x| < R} \; \vec{E}_c^{\,2}
\;d^3x\; \propto {\rm const} - \frac{\as(R)}{\pi} \frac{1}{R} \;.
\label{9}
\eeq
The pole mass assumes that all energy is
counted, i.e. $R\ra \infty$. Since in QCD the interaction becomes strong at
$R_0 \sim 1/\Lam$, the domain outside $R_0$ would yield an uncontrollable and 
physically senseless contribution to the mass $\sim \Lam$ \cite{gurman}.

Being a classical effect originating at a momentum scale well below $m_Q$, 
this uncertainty can be traced in the usual perturbation
theory, where it manifests itself in higher orders 
as a so-called $1/m_Q$ infrared (IR) renormalon singularity in the perturbative
series for the pole mass 
\cite{pole,bbpole}, see Fig.~2.

Nonetheless, the inclusive widths can be
theoretically calculated since they are governed, instead, by 
well-defined short-distance running masses $m_Q(\mu)$ with the 
Coulomb energy originating from distances $\gsim 1/\mu$ peeled off.  
It is precisely 
this short-distance running mass that can be
extracted from experiment with, in principle, unlimited accuracy: the pole mass
does not enter any genuine short-distance observable at the level of
nonperturbative corrections \cite{pole}. 

Applied to the inclusive widths, it suggests certain information
about the importance of higher order perturbative corrections: if masses
entering \eq{7} are the pole masses, the perturbative series 
\beq
\Gamma_{\rm sl}^{\rm pert} = \Gamma_0 \appa^{\rm pert}=\Gamma_0\left(1+
a_1 (\as/\pi) + a_2 \left(\as/\pi\right)^2 +...
\right)
\label{10}
\eeq
is poorly behaved,
with coefficients $a_k$ factorially growing, which makes  
the radiative correction factor uncalculable in principle with an 
accuracy $\sim \Lam/m_b$. In contrast, if one uses the
short-distance masses, the higher-order corrections become smaller and the
factor $\appa^{\rm pert}$ becomes calculable with the necessary 
precision \cite{pole,bbz}.

This seemingly academic observation, in reality proved to
underlie the pattern of the corrections from the very 
first terms. Remarkably, the actual model-independent calculations 
of $\Gamma_{\rm sl}$ through observables
measured in experiment are very stable against perturbative corrections.
Including ${\cal O}(\as^2)$ terms in the extraction of the $b$ pole mass from,
say, the $\rm e^+e^- \ra \bar b b$ threshold region \cite{volmb} noticeably
increases its value. However, the parallel perturbative
improvement in calculating the width yields an essential suppression of the
perturbative factor, $\appa$, so that the two effects offset each other almost
completely \cite{upset}.

This conspiracy is not unexpected: the appearance of large corrections at both
stages is an artefact of using the ill-defined pole mass in the
intermediate calculations. The situation is peculiar since the actual
nonperturbative effects appear only at the level $1/m_Q^2$, whereas the pole
mass is infrared ill-defined already at an accuracy of $1/m_Q$.
The failure to realize this fact led to the superficial
suggestion \cite{savage} that even in beauty particles the perturbative
corrections may go out of theoretical control; a more careful analysis
\cite{upset,bbbsl} showed that this is not the case.

Moreover, the OPE requires using short-distance running masses 
$m_{b,c}(\mu)$ normalized at $\mu \sim 1\GeV$ \cite{pole}. It has been done in 
\cite{upset} and
demonstrated that neither these masses nor the perturbative corrections to 
the width
$\appa$ show significant contributions from higher orders. 

To summarize, the idea that the perturbative corrections in the
extraction of $|V_{cb}|$ from $\Gamma_{\rm sl}(B)$ are large 
comes from an inconsistent usage of ill-defined pole masses:\\
\hspace*{2em}$\bullet$ It is 
`difficult to extract' accurately $m_b^{\rm pole}$ from 
experiment; in any
given calculation it is easy to identify the effects that were left out, 
which can
change its value by $\sim$ $200\MeV$. This uncertainty leads to a `theoretical
error' $\delta_I$ in $\Gamma_{\rm sl}(B)$ of $\sim$ $10\%$.\\
\hspace*{2em}$\bullet$ When routinely 
calculating $\Gamma_{\rm sl}(B)$ in terms of
the 
pole masses, there are significant higher order corrections $\delta_{II}
\approx 10\%$. 

The naive conclusion drawn from such experience \cite{neubmori} is
that one cannot reliably calculate the width without $\sim 20\%$ uncertainty:
$$
\delta 
\Gamma_{\rm sl}/\Gamma_{\rm sl}= \delta_I + \delta_{II} \simeq 20\%\;
\; \leftrightarrow \delta |V_{cb}|/|V_{cb}| \simeq 10\%\;\;.
$$
On the contrary, theory predicts a strong anticorrelation
between $\delta_I$ and
$\delta_{II}$ in a consistent perturbative 
calculation, and that was explicitly checked in
\cite{upset,bbbsl}. The net impact of the calculated (presumably
dominant) second-order ${\cal O}(\as^2)$ corrections on the value of $|V_{cb}|$
appeared to be less than $1\%\,$! Moreover, just neglecting {\em all}
perturbative corrections altogether, both in the semileptonic width and
in extracting $m_b$ from experiment, yields a $|V_{cb}|$ smaller by less than 
$5\%$ \cite{volpriv}. \vspace*{.2cm}

Recently, all-order corrections
associated with the running of $\as$ in one-loop diagrams (referred to as BLM 
approximation) were calculated in \cite{bbbsl}.
Using the most accurate model-independent determination of $m_b$ 
\cite{volmb}, one gets \cite{upset}
$$
|V_{cb}|=0.0413\left(\frac{{\rm BR}(B\rightarrow X_c\ell\nu)}{0.105}
\right)^{\frac{1}{2}}\left(\frac{1.6\,\rm ps}{\tau_B}\right)^{\frac{1}{2}}
\times
$$
\beq
\left(1-0.012\frac{(\tilde \mu_\pi^2-0.4\,\rm GeV^2)}{0.1\,\rm
GeV^2}\right)\cdot \left(1-0.006\frac{\delta m_b^*}{30\,\rm MeV}\right)\;.
\label{26b}
\end{equation}
The main source of theoretical uncertainty is the exact value of 
$\mu_\pi^2$ (marked with a tilde in \eq{26b}, indicating that a particular
field-theoretic definition is assumed), which enters
through the value of $m_b-m_c$, \eq{8}. 
A dedicated analysis of the lepton spectra will
reduce this uncertainty. At the moment a reasonable estimate of the 
uncertainty in $\mu_\pi^2$ is about $0.2\GeV^2$, leading to a 
$2.5\%$ uncertainty in $|V_{cb}|$.

The dependence on $m_b$ is minor; since we rely here on the
well-defined short-distance mass $m_b^*$, there is no intrinsic
uncertainty in it. The analysis \cite{volmb} estimated $\delta m_b^*
\simeq 30\MeV$; even for $\delta m_b^*
\simeq 60\MeV$,  
the related uncertainty in $|V_{cb}|$ is only $1.2\%$.

As previously explained, 
the actual impact of the known perturbative corrections when 
relating the
semileptonic width to other low-energy {\em observables} 
is very moderate, and there
is no reason to expect the higher-order effects to be significant. With the 
{\em a priori} dominant all-order BLM corrections calculated \cite{bbbsl}, one
may be concerned only with the true two-loop effects ${\cal O}(\as^2)$. 
These have not been calculated
completely yet; however, the recent ${\cal O}(\as^2)$ calculation \cite{czar} 
in the small velocity kinematics 
suggested that they must be 
small. There are some enhanced higher-order non-BLM corrections 
that are specific to 
the inclusive widths
\cite{ive}. They have been accounted for in the analyses \cite{vcb,upset}, but
went beyond those in \cite{bn,bbbsl}. 
Thus it seems
unlikely that as yet 
uncalculated second-order corrections can change the width by
more than $2$--$3\%$; therefore, 
assigning an additional uncertainty of $2\%$ in
$|V_{cb}|$ is a quite conservative estimate.

Adding up these uncertainties we arrive at 
\beq
\left(\delta|V_{cb}|/|V_{cb}|\right)\vert_{\rm th}\; \lsim \;5\%\;\;.
\label{17}
\eeq
The
theoretical accuracy in extracting of $|V_{cb}|$ appears to be better than its 
current
experimental counterpart. 
This method can be improved further in a 
model-independent way. I think that the $2\%$ level 
of a {\em defensible} theoretical precision can be ultimately reached here; 
an essential
improvement beyond that is questionable, because of effects of 
higher-order power corrections and
possible violations of duality.

Similarly, $|V_{ub}|$ is directly related to 
the total $b\ra u$ semileptonic width \cite{upset}:
\beq
|V_{ub}|=0.00458 \cdot \left[{\rm BR}(B\rightarrow X_u\ell\nu)/0.002
\right]^{\frac{1}{2}}\,\left(1.6\,\rm ps/\tau_B\right)^{\frac{1}{2}}\;.
\label{18}
\eeq
Recently, ALEPH announced \cite{aleph} a
model-independent measurement of the 
inclusive $b\ra u\, \ell\nu$ width: 
$ {\rm BR}(B\rightarrow X_u\ell\nu) = 0.0016\pm 0.0004 $.
I cannot judge the reliability of the quoted error bars in this 
sophisticated analysis; it certainly will be clarified soon.
Accepting this input literally, I arrive at the model-independent result
\beq
|V_{ub}|/|V_{cb}|\; = \; 0.098\pm 0.013\;\;.
\label{19}
\eeq
The theoretical uncertainty in converting $\Gamma(B\ra X_u\, \ell\nu)$ into
$|V_{ub}|$ is a few times smaller.

Let me briefly
comment on the literature. It is sometimes stated \cite{neubmori,neubert} 
that the uncertainty
in $\Gamma_{\rm sl}$ is at least $20\%$. The origin of such
claims is ignoring the subtleties related to 
using the pole mass in the
calculations and considering separately the perturbative corrections to the
pole masses, and to the widths expressed in terms of $m_Q^{\rm pole}$. This is
inconsistent on theoretical grounds \cite{pole}, whose relevance was 
confirmed by the
concrete numerical evaluations 
\cite{upset,bbbsl}. The
dependence on $m_b$ and $m_b-m_c$ used to determine the uncertainty in
$|V_{cb}|$ was calculated erroneously in \cite{neubmori} (cf. \cite{upset}), 
apparently because of an arithmetic mistake that led to a significant 
overestimate.
Finally, no argument was 
given to justify a sevenfold boosting of
the theoretical uncertainty in $m_b$ obtained in the dedicated analysis
\cite{volmb}.

\subsection{Exclusive zero recoil $B\ra D^*\,\ell \nu$ rate}

Good control of all QCD effects in $\Gamma_{\rm sl}$ was due
to the fact that removing constraints on the final state to which
decay partons can hadronize, makes such a probability a short-distance
quantity amenable to a direct OPE expansion. A similar approach to the 
exclusive zero-recoil decay rate $B\ra D^*\,\ell \nu$
yielded quite an accurate determination of $|V_{cb}|$ as well 
\cite{vcb,optical},
though with a more significant irreducible model dependence and a larger
intrinsic uncertainty. The limitation is twofold: constraining the
decays to a specific final state makes the transition not a genuinely
short-distance effect; it also suffers from 
a larger expansion parameter, namely $1/m_c$ vs. $1/m_b$. 

Near zero recoil the decay is
governed by the single hadronic formfactor $F_{D^*}$. In the infinite mass
limit $F_{D^*}=1$ holds; 
for finite $m_{b,c}$ it acquires corrections:
\beq
F_{D^*} = 1 -(1-\eta_A)+ \delta_{1/m^2} + ...\;\;.
\label{22}
\eeq
The effect of the nonperturbative domain starts with the terms 
$\sim 1/(m_c,m_b)^2$ \cite{vs,luke}, but
otherwise is rather arbitrary, depending on the details of the 
long-distance dynamics in the form of wavefunction overlap. This opened 
the field
for speculations and controversy \cite{neubpr}.

The situation as it existed by 1994 was summarized in reviews by 
Neubert \cite{neubtasi}:
\beq
\eta_A=0.986\pm 0.006 \qquad \qquad
\delta_{1/m^2} =(-2\pm 1)\% \;\;,
\label{23}
\eeq
yielding $F_{D^*}\simeq 0.97$, and was assigned the status of ``one of the most
important and, certainly, most precise predictions of HQET". Nowadays we
believe that the actual corrections to the symmetry limit are larger, and the
central theoretical value lies rather closer to $0.9$ \cite{vcb,optical}. 
Perturbative-wise, it has been pointed out \cite{comment} that the 
improvement \cite{neubimp} of
the original 
one-loop calculation was incorrect, and the proper estimate 
is $\eta_A\approx 0.965 \pm 0.025$; subsequent calculations of the 
higher-order BLM corrections \cite{bbbsl,flaw} confirmed it: 
$\eta_A\approx0.965 \pm 0.02$. The purely perturbative chapter was 
closed recently with the complete two-loop ${\cal O}(\as^2)$ result
\cite{czar} $\eta_A^{(\rm 2\, loop)} =0.960 \pm 0.007$;
however, the inherent irreducible uncertainty of the
{\em complete} perturbative series for $\eta_A$ 
exceeds the quoted one by a factor
of three \cite{bbbsl,ns,pert}.

If the mass of the charm quark were a few times larger, in practice
the two-loop calculation would have been the whole story for
$F_{D^*}$. In reality, the power corrections originating from the domain of
momenta below $\sim 0.6\GeV$ appear to be more significant. 
Not much can be said about them without model assumptions; 
they have been shown to be negative and
exceed about $0.04$ \cite{vcb,optical} in magnitude.

The idea of this dynamical approach was to consider the sum over all 
hadronic states in the zero recoil kinematics; 
such a rate sets an upper bound for the 
production of $D^*$. This {\em
inclusive} quantity is of a short-distance nature and can be calculated in QCD
using the OPE. The result through order $1/m^2$ is
\beq
|F_{D^*}|^2 +\sum_{\epsilon_i<\mu} |F_i|^2 = \xi_A(\mu) 
-\frac{\mu_G^2}{3m_c^2} -
\frac{\mu_\pi^2-\mu_G^2}{4}
\left(\frac{1}{m_c^2}+\frac{1}{m_b^2}+\frac{2}{3m_cm_b}
\right)\;\,,
\label{25}
\eeq
where $F_i$ are the transition formfactors to charm
states $i$ with the mass $M_i=M_{D^*}+\epsilon_i$, and $\xi_A$ is a
perturbative factor (the role of $\mu$ will be addressed later). Considering
a similar sum rule for another type of `weak current', say $\bar{c}
i\gamma_5 b\,$, yields 
\beq
\sum_{\tilde \epsilon_k<\mu} |\tilde F_k|^2 \;= \;
\left(\frac{1}{2m_c}- \frac{1}{2m_b}\right)^2\,
\left(\mu_\pi^2-\mu_G^2\right)
\label{26}
\eeq
with the tilde 
referring to the quantities occurring in the transitions induced by 
this hypothetical current. These sum rules (and similar ones at arbitrary
momentum transfer), established in \cite{vcb,optical}, have been subjected to 
a critical scrutiny for two years, but are now accepted and constitute 
the basis for currently used estimates of $F_{D^*}$.

Since \eq{26} is the sum of certain transition probabilities, it 
results in a rigorous lower bound
\beq
\mu_\pi^2\;>\;\mu_G^2 \simeq 0.4\GeV^2\;\;.
\label{27}
\eeq
The sum rule (\ref{25}) then leads to the model-independent lower bound for 
$\delta_{1/m^2}$: 
\beq
-\delta_{1/m^2}\;>\;\left(M_{B^*}^2-M_B^2\right)/8m_c^2\simeq 0.035\;.
\label{28}
\eeq
The actual estimate depends essentially on the value of $\mu_\pi^2$. It was
suggested in \cite{vcb} to estimate the contribution of the excited states in
the \lhs of the sum rule (\ref{25}) from $0$ to $100\%$  of the power
corrections in the right-hand side:
\beq
-\delta_{1/m^2}=(1+\chi)\left(\frac{M_{B^*}^2-M_B^2}{8m_c^2}
+\frac{\mu_\pi^2-\mu_G^2}{8}
\left(\frac{1}{m_c^2}+\frac{1}{m_b^2}+\frac{2}{3m_cm_b} \right)
\right),\;\; 0\le \chi \le 1\:.
\label{29}
\eeq
If so, one arrives at \cite{vcb}
$$
-\delta_{1/m^2}= (5.5\pm 1.8)\% \qquad \mbox{ at } \; \mu_\pi^2=0.4\GeV^2
$$
\beq
\,-\delta_{1/m^2}= (6.8\pm 2.3)\% \qquad \mbox{ at }
\mu_\pi^2=0.5\GeV^2\;\mbox{ }
\label{30}
\eeq
$$
-\delta_{1/m^2}= (8.1\pm 2.7)\% \qquad \mbox{ at } \; \mu_\pi^2=0.6\GeV^2
$$

The QM meaning  of the sum rules is transparent \cite{optical}. The
act of a semileptonic 
decay of the $b$ quark is its instantaneous replacement by a $c$
quark. In ordinary QM the overall probability of the produced state to
hadronize to something is exactly unity.
Why are there 
nonperturbative corrections in the sum rule? The answer is that
the `normalization' of the weak current $\bar c \gamma_\mu \gamma_5 b$ is not
exactly unity and depends, in particular, on the external gluon field.
Expressing the QCD current in terms of the nonrelativistic fields used in
QM one has, for example, through order $1/m^2$:
\beq
\bar c \gamma_k \gamma_5 b \leftrightarrow \sigma_k -
\left(\frac{1}{8m_c^2}(\vec\sigma
i\vec D)^2\sigma_k+\frac{1}{8m_b^2}\sigma_k(\vec\sigma
i\vec D)^2\;-
\frac{1}{4m_cm_b}(\vec\sigma i\vec D )\sigma_k(\vec\sigma i\vec D)
\right) .
\label{32}
\eeq
The last term just yields the correction seen in the \rhs of the
sum rule.
Let me note that in the standard HQET analysis, the first
two terms in the brackets are missing (see, e.g., Ref.~\cite{neubpr}) and
the dominant effect $\sim 1/m_c^2$ is lost; the nonrelativistic
expansion was correctly done in
the works by the Mainz group \cite{pirjol}.

The inequality $\mu_\pi^2 >\mu_G^2$ in QM expresses the positivity of the Pauli
Hamiltonian $\frac{1}{2m}(\vec\sigma\,i\vec D\,)^2 =
\frac{1}{2m}((i\vec D)^{\,2}-\frac{i}{2}\sigma G)$ \cite{volkin}.  It 
is interpreted as 
the Landau precession of a charged (colored) particle in the
(chromo)magnetic field where one has $\aver{p^2}\ge |\vec B|$. 
Although the QM
average of $\vec B$ in the 
$B$ meson is suppressed, the chromomagnetic field is proportional to the 
spin of 
the light degrees of freedom and
is thus essentially non-classical, which enhances the bound and makes up for
the suppression. 

The perturbative factor
$\xi_A$ is not equal to $\eta_A^2$ \cite{optical,pole} 
and depends on the separation scale
$\mu$. Unlike $\eta_A$, which in principle cannot be defined
theoretically with better than a few percent accuracy, $\xi_A(\mu)$ is 
well-defined; 
no significant uncertainty is associated with it, 
$\xi_A \simeq (0.99)^2$.

Allowing a very moderate
variation of $\mu_\pi^2$ between $0.4\GeV^2$ and
$0.6\GeV^2$ only, we see that $-\delta_{1/m^2}$ varies between 
$3.5\%$ and $11\%$;
moreover, since there are no model-independent arguments to prefer any
part of the interval, the whole range must be considered equally
possible. Adding small perturbative corrections we end up with the 
reasonable estimate $F_{D^*}\approx 0.9\,$.
It is curious to note that at a `central' value $\chi=0.5$
the dependence of the zero-recoil decay rate on $\mu_\pi^2$ 
through $\delta_{1/m^2}$ effect
practically coincides with that of $\Gamma_{\rm sl}(B)$ (see \eq{26b}) although
they actually vary in opposite directions. 
The typical size of the $1/m_Q^2$ corrections to the exclusive zero-recoil
decay rate is thus significant, around $15\%$, which 
is expected since they are 
driven by the scale $m_c\simeq 1.3\GeV$. It is evident that $1/m_c^3$
corrections in $F_{D^*}$ not addressed so far are {\em at least} about
$\frac{1}{2}(0.15)^{3/2} \simeq 2$--$3\%$.\footnote{This is consistent 
with the fact that the $1/m_Q^3$ IR
renormalon ambiguity in $\eta_A^2$ constitutes $5\%$ at $\Lambda_{\rm QCD}^{\rm
\overline{MS}}\simeq
220\MeV$ \cite{pert}.} 

Thus, I believe that the current theoretical technologies do not allow to
reliably predict the zero recoil formfactor $F_{D^*}$ with a precision 
better than
$5$--$7\%$ in a model-independent way; its value is expected to be
approximately $0.9$, although a correction to the symmetry 
limit twice smaller, as  
well as larger deviations, are possible. It is encouraging that the `educated
guess' $F_{D^*}\simeq 0.9$, which emerged 
from the first -- and so far the only -- 
dynamical QCD-based consideration
\cite{vcb,optical}, yielded a  value of $|V_{cb}|$ close
to a less uncertain result obtained from $\Gamma_{\rm sl}(B)$.

Future, more accurate data will enable us to measure $F_{D^*}$
with a theoretically informative precision using $|V_{cb}|$ from $\Gamma_{\rm
sl}(B)$, and thus provide us with deeper insights into the dynamics of strong
forces in the heavy quark system. 

Certain statements in the literature deserve comments. 
Neubert claimed \cite{update,ns} that the sum rule (\ref{25})
cannot be correct, since $1/m^2$ renormalons allegedly mismatch in it. It was
failed to realized in these papers 
that in Wilson's OPE the IR renormalons are always
absent from any particular term. On the other hand, IR renormalon calculus can 
still
be applied if the OPE relation is considered in the pure perturbation theory
itself, and
formally setting $\mu=0$.  
In particular, this amounts to subtracting a `perturbative piece' from
the observable probabilities.
However, in this way the
perturbative terms appear in the left-hand side as well, and these
terms were ignored in Refs.~\cite{update,ns}.

It was suggested in Ref.~\cite{lig} that the higher-order radiative
corrections to the sum rules are too large and allegedly make them next to
useless. Such conclusions emerged from
a theoretically inappropriate treatment. 
The concrete numerical analysis in the OPE quoted
below, on the contrary, 
suggests a quite moderate impact of radiative corrections. 
According to \cite{lig}, the perturbative corrections to the sum rule 
of the type of \eq{26} weaken 
the bound for the expectation value of the kinetic operator 
to such an extent that it becomes non-informative. 
One must realize that, in the quantum field theory, the renormalized operators
can be defined in different non-equivalent ways; $-\lambda_1$ addressed in 
\cite{lig} is
known to be different from $\mu_\pi^2$. 
Moreover, the only field-theoretic
definition of the kinetic operator $\bar{Q} (i\vec D\,)^2 Q$ given so far was 
made in \cite{optical} and, for it, $\mu_\pi^2>\mu_G^2$ always holds.
As for $-\lambda_1$, a parameter in HQET, its definition 
beyond the classical level has
never been given; the procedure adopted in Ref.~\cite{lig} reduces to 
an attempt
to completely subtract the `perturbative piece' of $\mu_\pi^2(\mu)$:
\beq
-\lambda_1= \mu_\pi^2(\mu)-c_1\left(\as(\mu)/\pi\right)\mu^2-
c_2\left(\as(\mu)/\pi\right)^2\mu^2 - ...
\label{34}
\eeq
(the method to calculate $c_i$ was elaborated in \cite{optical}). Yet it has
been known for a long time \cite{fail} that such a program theoretically 
cannot be performed: the series in \eq{34} is factorially divergent and cannot
be assigned a meaningful number. 
No wonder the second-order BLM correction
calculated in \cite{lig} seemed to be dangerously large: there can be
no bound established for a quantity that is not defined. Moreover, the 
situation is clear in the BLM
approximation, where all $c_i$ are readily calculated: the series, whose second
term was discussed in \cite{lig}, is divergent and sign-varying, so
using merely the second term is misleading for any numerical estimate.

The above subtleties are peculiar to the field-theory analysis. 
Inequality $\mu_\pi^2>\mu_G^2$  must hold in {\em any} 
QM model relying on a potential description without additional 
degrees of freedom, if the
heavy quark Hamiltonian is consistent with QCD. Unfortunately, a failure to
realize this fact is seen in a number of recent analyses.

A second-order BLM analysis of the sum rule (\ref{25}) for $F_{D^*}$
was also attempted in \cite{lig} and claimed to destroy its predictive power 
(the first-order calculation had been performed in 
\cite{optical}). This calculation, however, was not done 
consistently, and the
actual effect is smaller \cite{pert}.
Let me define 
$\eta_A(\mu)\equiv \xi_A(\mu)^{1/2}$;
the quantity $\eta_A(\mu)$ must be added to $\delta_{1/m^2}$ in the framework
of the OPE instead of $\eta_A$ in the model
calculations. Then, at a reasonable choice
$\mu\simeq 0.5\GeV$, $\Lam^{(V)}=300\MeV$, one has
\begin{eqnarray}
\eta_A(\mu)\; = & 1  &  \;\;\;\mbox{ tree level}\\
\eta_A(\mu)\; = & 0.975 &   \;\;\;\mbox{ one loop}\\
\eta_A(\mu)\; = & 0.99  &  \;\;\;\mbox{ all-order BLM}
\label{35}
\end{eqnarray}
Clearly, the effect of the calculated perturbative corrections is not 
drastic
and $\eta_A(\mu)$ is very close to the value of $0.98$ adopted in the original 
analysis \cite{vcb}.

Conceptually, the deficiency of the alternative application of the original sum
rules of Refs.~\cite{vcb,optical} adopted in \cite{lig}, is to gauge $\xi_A$ on
the value of $\eta_A^2$ as it has been defined in the HQET (the idea that
$\xi_A$ is to be identified with $\eta_A^2$ ascends to \cite{ns,update}). 
However, it is 
$\eta_A$ that is ill-defined, and only for this reason must the difference
between the stable Wilson coefficient $\xi_A$ and $\eta_A^2$ suffer from
large corrections. It is worth noting that, in reality, $\eta_A$
cannot be equal to a matching coefficient of $\bar c \gamma_\mu \gamma_5
b$ to a corresponding current in {\em any} effective field theory.

Smaller theoretical uncertainty,
$2.5\%$ and $3\%$, 
is now quoted by Neubert for $\delta_{1/m^2}$ and $F_{D^*}$, 
respectively. The
former was obtained in Ref.~\cite{update}, 
in what he calls a ``hybrid approach'', which reduces to
assigning the fixed value $\mu_\pi^2=0.4\GeV^2$ and using
it in the sum rule (\ref{25}) within the same model assumption of \eq{29}
as suggested in \cite{vcb}: 
$0\le \chi\le 1$.\footnote{I disagree with the statements of \cite{update},
reiterated in later papers, suggesting that the original analysis
\cite{vcb,optical} missed some elements of the heavy quark spin-flavor symmetry;
on the contrary, it was stated in the latter paper that all these relations
automatically emerge from the sum rules that replace the QM wavefunction 
description in the quantum field theory.} Correspondingly, the quoted number
for $\delta_{1/m^2}$ practically 
coincided with the first line of \eqs{30}. In reality, allowing 
$\mu_\pi^2$ to vary within any reasonable interval 
significantly stretches the uncertainty.
Moreover, the analysis \cite{update} was based on using $\eta_A$ as a
perturbative factor assuming, literally, that in the proper treatment the final
result would not be changed numerically -- which is just the case according to
Ref.~\cite{lig}. On top of that, the uncertainty in the definition of
$\eta_A$ due to $1/m^2$ and $1/m^3$ IR renormalons constitutes 
$2$--$3\%$ each and
is an additional one in the adopted usage of the sum rules. Altogether, the 
stated theoretical
confidence of those estimates cannot be accepted as realistic.

Recently, 
Ref.~\cite{capneub} claimed to have established an intriguing relation
between the slope and the curvature of the formfactor near the zero recoil
point, using analyticity and unitarity of 
the amplitudes. If correct, this would reduce the experimental
uncertainties in extrapolating the rate to the zero recoil. However, both
the below-threshold contribution to the dispersion integral 
and the $1/m_c$ 
power corrections \cite{vain} to the heavy quark
symmetry relations were grossly underestimated; therefore the
relation stated in \cite{capneub} rather should not be used for deriving 
model-independent experimental results.

\section{The semileptonic branching fraction}

The QCD-based HQE provides a systematic framework for calculating the total
widths of heavy flavors, which are not amenable to the traditional methods of
HQET. The difference between nonleptonic and semileptonic widths appears only 
at a quantitative level. The only assumption is that the mass of a decaying 
quark (actually, the energy release) is sufficiently large;  for a review, see 
\cite{stone}.
 
The overall semileptonic branching ratio ${\rm BR_{sl}}(B)$ seems to be of a
particular practical interest: while the simple-minded parton estimates yield
${\rm BR_{sl}}(B)\simeq 15\%$ \cite{parton}, experiments give smaller
values ${\rm BR_{sl}}(B)\simeq 10.5$--$11.5\%$. 
The leading $1/m_Q^2$ effects in the nonleptonic widths were calculated
in \cite{buv,bs,bbsuv}; they exhibit some cancellations and one literally gets
\cite{buffling} a downward shift $\lsim 0.5\%$.
Estimated $1/m_b^3$ corrections do not produce a significant effect either 
\cite{mirage}. As a result, most of the attention was paid to a more accurate
treatment of the perturbative corrections, including  
the effect of the charm mass in the final state. It was found
\cite{bagan,volbr} that the nonleptonic width is indeed boosted up.

Since the inclusive widths are expanded in inverse powers of 
energy release, one expects larger corrections or even a
breakdown of the expansion and violation of duality in the channel $b\ra c\bar 
c s(d)$; however, this channel can be isolated via charm counting \cite{buv}. 
The original 
experimental estimate $n_c \lsim 1.15$ did not allow one 
to attribute the apparent
discrepancy to it, and gave rise to the so-called `${\rm
BR_{sl}}$ versus $n_c$' problem.

The perturbative corrections in the $b\ra c\bar u d$ itself cannot naturally 
drive $\rm BR_{sl}$ below $12.5\%$; the calculation 
for the $b\ra c\bar c s(d)$
channel is less certain and, in principle, admits increasing the width by a
factor of $1.5$--$2$, leading to $n_c\simeq 1.25$--$1.3$. In the latter
case a value of $\rm BR_{sl}$ as low as $11.5\%$ can be accommodated.

The experimental situation with $n_c$ does not seem to be quite settled yet:
$n_c =1.134\pm 0.043$ (CLEO), $n_c =1.23\pm 0.07$ (ALEPH).
In
a recent analysis \cite{dunietz} it was argued that consistency requires a
major portion of the final states in $b\ra c\bar c s$ to appear as 
modes with kaons but $D_s$, which previously escaped proper 
attention.  
This allows for a larger value $n_c\simeq 1.3$ needed to
resolve the problem with $\rm BR_{sl}$. The dedicated theoretical analysis
\cite{bsu} shows that, indeed, the dominance of such modes is natural and 
does not require violation of duality. Thus, if 
a larger value of $n_c\simeq 1.25$ is confirmed experimentally, 
the problem of $\rm BR_{sl}$
will not remain.

In my opinion, however, we cannot consider even this successful scenario  as a
complete QCD-based theoretical prediction of $\rm BR_{sl}$; a strong
enhancement of a tree-level-unsuppressed channel raises doubts about the
trustworthiness of its one-loop calculations. The possibility to get 
the necessary enhancement should be rather viewed as an indication of the 
presence of large effects working in the right direction.

\section{Lifetimes of beauty particles}

A thoughtful application of the HQE to charm lifetimes demonstrated
that the actual expansion parameter appeared to be too low to ensure a 
trustworthy accurate description, so that {\em a priori} one expects only 
emergence of the 
qualitative features. Surprisingly, in most cases the expansion
works well enough even numerically (for a recent review, see
\cite{bigdor}).

Applying the expansion to beauty particles one expects a decent
numerical accuracy, although the overall scale of the effects is
predicted to be small, making a challenge to experiment: 
\beq
\begin{array}{cclcl}
\tau_{B^-}/\tau_{B^0} & \simeq & 1+0.04
\left(\frac{f_{B}}{180\,\rm MeV}\right)^2 &\cite{mirage} &
\qquad \mbox{ EXP: } \qquad 1.04\pm 0.04\\

\overline \tau_{B_s}/\tau_{B^0} & \simeq & 1+{\cal O}(1\%) &
\cite{mirage} & \qquad \mbox{ EXP: } \qquad 0.97\pm 0.05\\

(\tau_{B_s^L}-\tau_{B_s^S})/\tau_{B_s} & \simeq & 0.18 \left(
\frac{f_{B_s}}{200\,\rm MeV}\right)^2 & \cite{vsku} & \\

\tau_{\Lambda_b}/\tau_{B^0} & \approx &  0.9 & &
\qquad \mbox{ EXP: } \qquad 0.78\pm 0.06 \\
\end{array}
\label{lam}       
\eeq
These differences appear mainly as $1/m_b^3$ corrections and, depending on 
certain
four-fermion matrix elements, cannot be predicted at present very accurately,
in particular in baryons. For mesons
the estimates are based on the vacuum saturation approximation, which 
cannot be
exact either. The impact of non-factorizable terms has been studied a 
few years ago
in \cite{WADs} and possibilities to directly measure the matrix elements in
future experiments were suggested.

The apparent agreement with experiment is obscured by reported lower values of
$\tau_{\Lambda_b}$. Since the baryonic matrix elements
are rather uncertain, a few model estimates have been done \cite{Rosner}. 
All seem to fall short; 
however, this might be attributed to deficiencies of the simple quark model.
Nevertheless, it was
shown \cite{boost} that irrespective of the
details one cannot have an effect exceeding $10$--$12\%$ while residing in the
domain of validity of the standard $1/m_Q$ expansion itself; the natural
`maximal' effects that can be accommodated are $\sim 7\%$ and $\sim 3\%$
for weak scattering and interference,  respectively.

Thus, if the low experimental value of $\tau_{\Lambda_b}$ 
is confirmed,  it will require a certain
revision of the standard picture of the heavy hadrons and of convergence of the 
$1/m_Q$
expansion for nonleptonic widths with the offset of duality in beauty
particles.

Recently, the problem of the accuracy of the calculations of $\delta \tau_B$
based on the factorization was emphasized again in \cite{ns2}. It is difficult
to agree, however, with the wide intervals, up to $\pm 20\%$ allowed for the
difference between $\tau_{B^+}$ and $\tau_{B^0}$; the constraints discussed in
\cite{WADs,boost} were missed. One can see that the values
of hadronic parameters saturating such large differences would move one beyond
the domain of applicability of the whole expansion used in \cite{ns2}.

\section{$1/m_Q$ expansion and duality violation}

Duality violation attracts more and more attention in the
context of the heavy quark theory; a recent extensive discussion was given in
\cite{inst}. The expansion in $1/m_Q$ is asymptotic. There are basically two
questions one can ask here: what is the onset of 
duality, i.e. {\em when} does the
expansion start to work? The most straightforward approach was first 
undertaken in
\cite{bm}, and no apparent indication toward an increased energy scale was
found. Another question, of {\em how} is the equality of the QCD
parton-based predictions with the actual decay rates achieved, was rarely
addressed.
An example of such a problem is easy to
give.

The
OPE states that no terms $\sim 1/m_Q$ can be in the widths and the leading
terms start with $1/m_Q^2$. However, the OPE {\em per se} cannot forbid a
scenario where, for instance,
\beq
\delta \Gamma_{H_Q}/\Gamma_{H_Q} \;\sim\ \;C \:
\sin{(m_Q \rho)}/(m_Q\rho)
\;,\;\;\; \rho \sim \Lam^{-1}
\;.
\label{60}
\eeq
In the actual strong
interaction, $m_b$ and $m_c$ are fixed, so 
from the practical viewpoint these types of corrections are 
not too different -- but the difference is profound theory-wise! It is a
specific feature of the OPE in Minkowski space, and it can hardly be addressed,
for example, in lattice calculations. Their complete control requires a deeper
understanding of the underlying QCD dynamics beyond the knowledge of first few
nonperturbative condensates.

The literal corrections of the type of \eq{60} are hardly possible; the
power of $1/m_Q$ in realistic scenarios is larger, and they 
must be eventually exponentially suppressed though, probably, 
starting at a higher 
scale \cite{inst}. But a theory of such effects is still in its embryonic
stage and needs an experimental input as well.

The possibility has been discussed for some time \cite{alt} 
to have certain unidentified corrections to the (nonleptonic) widths, 
eventually leading to the dependence
\beq
\Gamma^{\rm nl}_{H_Q} \sim M_{H_Q}^5
\label{61}
\eeq
and manifesting an explicit $1/m_Q$ effect:
\beq
\delta \Gamma_{H_Q}/\Gamma_{H_Q}\; \simeq\; \gamma \;\delta
M_{H_Q}/M_{H_Q}\;,\;\;\;\gamma\simeq 5\;.
\label{62}
\eeq
Such scaling in the intermediate energy domain cannot
literally contradict OPE if the offset of duality has not been passed yet. But
is this possibility natural?

Leaving aside the QCD-based arguments completely, one must still account for
the charm mass in the final state, and thus differentiate between $M_D$ and
$M_{\Lambda_c}$ in the 
decays of $B$ and $\Lambda_b$, respectively.
Although $M_{\Lambda_b}$ is notably larger than $M_B$, $M_{\Lambda_c}$ exceeds
$M_D$ by almost the same amount! Counting only the 
phase space factors in analogy
with the free quark decay one would get for the $b\ra c$ transitions
$\gamma \lsim 2$ \cite{upset}, which thus seems to be a more natural value in 
the `poor man on the street' hypothesis \cite{mart96} considered 
in \cite{alt}.
It is worth noting that the fit of $\gamma$ in charm particles does not 
convincingly indicate
favoring $\gamma=5$: incorporating the calculated $1/m_c^2$ and
$1/m_c^3$ corrections destroys it -- while discarding them as a part
of the `poor man' philosophy makes it impossible to explain the very different
values of $\tau_{D}$ and ${\rm BR_{sl}}(D)$. The short 
$\Lambda_b$ lifetime thus still seems to constitute an important problem 
for the whole heavy quark theoretical community.
\vspace*{0.25cm}\\

\noindent {\large \bf Acknowledgements} 
\vspace*{0.15cm}\\
\noindent
I am grateful to F.~Ferroni, G.~Martinelli
and other organizers of Beauty `96 for the opportunity to participate in this
meeting. I benefited from invaluable collaboration
with I.~Bigi, M.~Shifman and 
A.~Vainshtein. I am thankful to M.~Voloshin for many clarifications.
Informal discussions with V.~Braun, G.~Martinelli and C.~Sachrajda were
enjoyable, although short. It is a 
pleasure to thank P.~Schlein for detailed 
comments on the manuscript.
This work was supported in part by the NSF under the grant
number PHY 92-13313.

\end{document}